\documentclass{ws-procs10x7}
\usepackage{balance}
\usepackage{epsfig}
\newcolumntype{d}[1]{D{.}{.}{#1}}

\def\Journal#1#2#3#4{{\it #1} {\bf #2}, #3 (#4)}

\makeindex
\begin{document}

\title{SEARCH FOR NEW PHYSICS IN RARE D DECAYS}
\author{Svjetlana  Fajfer and Sasa Prelovsek}

\address{Department of Physics, University of Ljubljana,
Jadranska 19, 1000 Ljubljana and \\
J. Stefan Institute, Jamova 39, P. O. Box 300, 1001 Ljubljana,
Slovenia\\ E-mail: svjetlana.fajfer@ijs.si, sasa.prelovsek@ijs.si\\
$~$\\
Talk given by S. Fajfer at International Conference On High Energy Physics (ICHEP 06)\\
 26 Jul-2 Aug 2006, Moscow, Russia}

\twocolumn[\maketitle\abstract{In many extensions of the standard model 
an additional up-like heavy quark appears. Its appearance 
induces flavour changing 
neutral current transition in the up-quark region at the tree level.  
 We  investigate possible effects of these models in the $c \to u l^+l^-$ 
transitions. 
First we determine impact of new physics on the relevant Wilson coefficients 
 and then we reevaluate the standard model long-distance 
contributions. 
We calculate  differential branching ratio for 
the $D^+ \to \pi^+ l^+ l^-$ and $D^0 \to \rho^0 l^+ l^-$ decays. 
Among all D decay modes these two are the simplest ones 
for the experimental studies. 
We also determine the forward-backward 
asymmetry for the $D^0 \to \rho^0 l^+ l^-$
 decay and 
we comment on the effects 
of the Littlest Higgs model in both decay modes.}

]


At low-energies new physics is usually 
expected in the down-like quark sector. Numerous studies of new  physics 
effects  were performed in the $s \to d$, $b \to s (d)$, 
$\bar s d \leftrightarrow \bar d s$, $\bar b d \leftrightarrow \bar d b$ and 
$\bar b s \leftrightarrow \bar s b$ transitions.

However, searches for new physics in the up-like quark sector at low energies 
were not so  
attractive. 
Reasons are following: a) flavour changing neutral current processes 
at loop level in the standard model suffer from the  
 GIM cancellation leading to very small effects in the $c \to u$ 
transitions. 
 The GIM mechanism acts in many extensions of the standard model 
too, making contributions of new physics insignificant.  
b)   Most of the charm meson processes,  where 
$c \to u$  and $c \bar u \leftrightarrow \bar c u$ transitions might occur  
are  dominated by the standard model long-distance  contributions
 \cite{burdman1} - \cite{bigi0}.

On the experimental side there are many studies of rare charm meson decays. 
The first  observed rare $D$ meson decay was  the radiative
 weak decay 
$D \to \phi \gamma$. Its rate $BR(D \to \phi \gamma)= 2.6^{+0.7}_{-0.6} \times 10^{-5}$ has been 
measured by Belle collaboration 
\cite{Belle1}  and hopefully other
 radiative weak charm decays will be observed soon\cite{CLEO_pll}. 

\vspace{0.2cm}


In the standard model (SM) \cite{burdman1} the contribution coming from the penguin 
diagrams in 
$\rm c\to u\gamma$ transition gives 
branching ratio of order $10^{-18}$. 
The QCD corrected
effective Lagrangian \cite{greub} gives $\rm BR(c\to u\gamma)\simeq3\times10^{-8}$. 
A variety of models beyond SM  were
investigated and it was found that the gluino exchange diagrams
\cite{sasa} within general minimal supersymmetric SM (MSSM) might lead to  the
enhancement
\begin{equation}
\rm\frac{BR(c\to u\gamma)_{{MSSM}}}{BR(c\to u\gamma)_{{SM}}} 
\simeq10^2.
\label{1}
\end{equation}



The inclusive $c\to u l^+l^-$ calculated at one-loop level in SM 
 \cite{prelovsek3}
was found to be suppressed by QCD corrections \cite{burdman2}. 
The inclusion of the renormalization group equations  
for the Wilson coefficients 
gave an additional significant 
suppression \cite{jure} leading to the rates  
$\Gamma(c\to ue^+e^-)/\Gamma_{D^0}=2.4\times 10^{-10}$ and
$\Gamma(c\to u\mu^+\mu^-)/\Gamma_{D^0}=0.5\times 10^{-10}$.   
These transitions are largely driven by virtual photon at low dilepton mass $m_{ll}$.

The leading 
contribution to $c\to ul^+l^-$ in general MSSM with conserved R parity 
comes from the one-loop diagram with 
gluino and squarks in the loop \cite{burdman2,prelovsek3,sasa}. 
It proceeds via virtual photon  
and significantly enhances the $c\to ul^+l^-$ 
spectrum at small dilepton mass $m_{ll}$. 
The authors of Ref. \cite{burdman2} have investigated supersymmetric 
(SUSY) extension of the SM with R parity breaking and they 
found that it can modify the rate. Using most recent CLEO \cite{CLEO_pll} 
results for the $D^+ \to \pi^+ \mu^+ \mu^-$ one can set the bound for the product of the 
relevant parameters entering 
the R parity violating $\tilde \lambda'_{22k} \tilde \lambda'_{21k} \simeq 0.001 $ 
(assuming that the 
mass of squark $M_{\tilde D_k} \simeq 100$ GeV). This bound gives the rates 
$BR_R(c\to ue^+e^-) \simeq1.6 \times 10^{-8}$ and  
$BR_R(c\to u \mu^+\mu^-) \simeq1.8 \times 10^{-8}$. 

Some  of models of new physics (NP) contain un extra up-like 
heavy quark 
inducing the flavour changing neutral currents at tree 
level for the up-quark sector \cite{FP-LH,barger,lang,abel,higuchi}. 
The isospin component of the weak neutral current is given in \cite{FP-LH} as
\begin{equation}
J_{W^3}^\mu = \frac{1}{2} \bar U_L^m \gamma^\mu \Omega U_L^m -  
\frac{1}{2} \bar D_L^m \gamma^\mu  D_L^m
\label{e2}
\end{equation}
with $L=\tfrac{1}{2}(1- \gamma_5)$ and mass eigenstates 
$U_L^m= (u_L,c_L,t_L,T_L)^T$, $D_L^m=(d_L,s_L,b_L)^T$.
The neutral current for the down-like quarks is the same as in 
the SM, while there are tree-level flavour changing transitions 
between up-quarks if $\Omega \not =I$. The elements of $4\times 4$ matrix 
$\Omega$ can be constrained by CKM  unitarity violations 
currently allowed by experimental data. Even more stringent bound 
on $c u Z$ coupling $\Omega_{uc}$ comes from the present 
bound on $\Delta m $ in $ D^0 - \bar  D^0$ transition. 
It gives $|\Omega_{uc}| \leq 0.0004$ and we use the upper bound to 
determine the maximal effect on rare $D$ decays in what follows.
In this case the dilepton mass distribution of 
the $c \to u l^+l^-$  differential branching ratio can be enhanced 
by two orders of magnitude in comparison with SM (see Fig.1). 

A particular version of the model with  
tree-level up-quark FCNC transitions is the Littlest Higgs model \cite{lee}.  
 In this case the  magnitude of the relevant $c \to u Z$ coupling  
$\Omega_{cu} =|V_{ub}||V_{cb}|v^2/f^2 \leq 10^{-5}$ 
is even further 
constrained via the scale $f\geq {\cal O}(1~{\rm TeV})$ 
by the precision electro-weak data. The smallness of $\Omega_{uc}$
 implies that the effect of this particular 
model on $c\to ul^+l^-$ decay and relevant rare $D$ decays is insignificant
 \cite{FP-LH}. 



The study of exclusive D meson rare decay modes is very difficult due to the 
dominance of the long distance effects \cite{burdman1} - \cite{prelovsek2}.
The inclusive $c \to u l^+ l^-$ can be tested in the rare decays 
$D \to \mu^+ \mu^-$, 
$D \to P (V) l^+ l^-$ \cite{burdman2,prelovsek3,burdman3}.

The branching ratio for the rare decay $D\to \mu^+ \mu^-$ 
is very small in the SM. 
The detailed treatment of this decay rate \cite{burdman2} 
gives $Br(D \to \mu^+ \mu^-) \simeq 3\times 10^{-13}$ \cite{burdman2}. This decay rate 
can be enhanced within a study which considers 
 SUSY with R parity breaking effects \cite{burdman2,bigi0}. 
Using the bound $\tilde \lambda'_{22k} \tilde \lambda'_{21k} \simeq 0.001 $ 
one obtains the limit $Br(D \to \mu^+ \mu^-)_R\simeq 4\times 10^{-7}$. 

The $D \to P (V) l^+ l^-$ decays offer another possibility to study the $c \to u l^+ l^-$ transition in charm sector. 
The most appropriate decay modes for the experimental searches 
are $D^+\to \pi^+ l^+l^-$ and  
$D^0\to \rho^0e^+e^-$.  In the following we present the possible maximal 
effect on these decays coming from general model with tree level 
$cuZ$ coupling at its upper bound $|\Omega_{uc}|=0.0004$. 
We already pointed out that in Littlest Higgs model, 
which is a particular version of these models, 
 the coupling $\Omega_{uc}$ is constrained to be 
smaller and the effects on rare $D$ decays are insignificant \cite{FP-LH}.

The    calculations of the long distance contributions in the  decays 
$D^+\to \pi^+l^+l^-$ and $D^0 \to \rho^0 l^+ l^-$ are presented 
in Refs. \cite{FP-LH,prelovsek2,prelovsek3}.
The contributions of the 
intermediate vector resonances $V_0=\rho^0,\omega,\phi$ with $V_0\to l^+l^-$ 
constitute 
an important long-distance contribution to the hadronic decay, 
which may shadow interesting 
short-distance contribution induced by $c\to ul^+l^-$ transition. 

Our determination of short and long distance contributions to 
$D^+\to \pi^+l^+l^-$ takes advantage of the available 
experimental data \cite{FP-LH}. 
This is a fortunate circumstance for this particular decay since the 
analogous experimental input is not available for determination of  
the other $D\to X l^+l^-$ rates in a similar way. The rate 
resulting from the amplitudes (14) and (19) of \cite{FP-LH} with   
$|\Omega_{uc}|=0.0004$ are given  in Figure 2 and Table 1.

We are unable to determine the  amplitude of 
the long-distance contribution to $D^0\to \rho^0 V_0\to \rho^0 l^+l^-$ 
using the measured rates for 
$D^0\to \rho^0 V_0$ since only  the rate of $D^0\to \rho^0 \phi$ is known 
experimentally. We are forced to use a model \cite{prelovsek2}, 
developed to describe all $D\to Vl^+l^-$ and $D\to V\gamma$ decays, 
and the resulting rates are presented in Figure 3 and Table 1.

\begin{figure}[b]
\epsfig{file=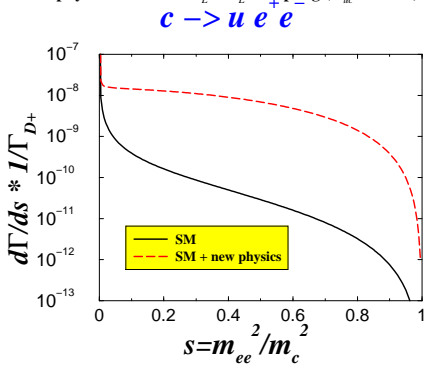,width=2.4in}
\caption{    The dilepton mass distribution 
$dBr/dm_{ee}^2$ for the inclusive decay 
$c\to u l^+l^-$   
as a function of the  dilepton mass square $m_{ee}^2=(p_++p_-)^2$.
}
\end{figure}

Therefore, the total rates for $D \to X l^+ l^-$ are dominated by the 
long distance resonant contributions at dilepton mass 
$m_{ll}=m_\rho,~m_\omega,~m_\phi$ 
and even the largest contributions from new physics are not expected to 
affect the total rate significantly \cite{burdman2,prelovsek3}. 
New physics could only modify the SM 
differential spectrum at low $m_{ll}$ below 
$\rho$ or spectrum at high $m_{ll}$ above $\phi$. 
 In the case of $D\to\pi l^+l^-$ differential decay 
distribution there is a broad  
 region at high $m_{ll}$ (see Fig. 2), 
 which presents a unique possibility to 
study $c\to ul^+l^-$ transition \cite{prelovsek3,FP-LH}.

\begin{table*}
\begin{center}
\tbl{Branching ratios for decays in which $c\to ul^+l^-$ transition 
can be probed. 
\label{tab1}}
{\begin{tabular}{@{}ccccccc@{}}
\toprule
 {\bf Br} & \multicolumn{2}{c|}{short distance } & total rate $\simeq$ & experiment\\
 & \multicolumn{2}{c|}{contribution only } & long distance contr. & \\  
\hline 
 & SM & SM + NP &    &   \\
\hline
$D^+\to \pi^+ e^+e^-$ & $6\times 10^{-12}$ & $8\times 10^{-9}$ & $1.9\times 10^{-6}$ & $<7.4\times 10^{-6}$\\
$D^+\to \pi^+ \mu^+\mu^-$ &  $6\times 10^{-12}$ & $8\times 10^{-9}$ & 
$1.9\times 10^{-6}$ & $<8.8\times 10^{-6}$\\
\hline
$D^0\to \rho^0 e^+e^-$ & negligible &$5\times 10^{-10}$ & $1.6\times 10^{-7}$
&$<1.0\times 10^{-4}$\\
$D^0\to \rho^0 \mu^+\mu^-$ & negligible &$5\times 10^{-10}$ & $1.5\times 10^{-7}$ & $<2.2\times 10^{-5}$\\
\botrule
\end{tabular}}
\end{center}
\end{table*}

The non-zero forward-backward asymmetry in $D\to \rho l^+l^-$ decay  arises 
only when $C_{10}\not =0$ (assuming $m_l\to 0$).  
The enhancement of the $C_{10}$ in the NP  models \cite{FP-LH} 
 is due to the tree-level $\bar u_L\gamma_\mu c_LZ^\mu$ coupling
 and leads to  nonzero asymmetry $A_{FB}(m_{ll}^2)$ shown in 
Fig. 4.   
The forward-backward asymmetry for  $D^0 \to \rho^0 l^+ l^-$ vanishes in 
SM ($C_{10}\simeq 0$), while it is reaching ${\cal O}(10^{-2})$  
in NP model with the extra up-like quark as shown in Fig. 4. 
Such asymmetry is still small and  difficult to be seen in 
the present or planned experiments given that the rate itself is already 
small.

\begin{figure}[b]
\epsfig{file=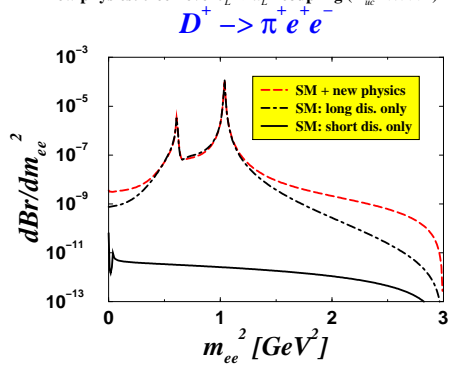,width=2.4in}
\caption{The dilepton mass distribution 
 $dBr/dm_{ee}^2$ for $D^+\to \pi^+e^+e^-$.   }
\label{fig2}
\end{figure}
\begin{figure}[b]
\epsfig{file=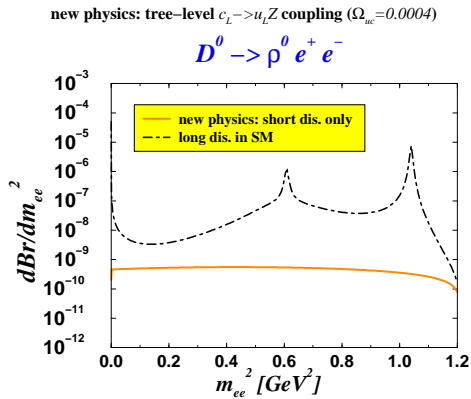,width=2.4in}
\caption{ 
The dilepton mass distribution for $D^0\to \rho^0e^+e^-$.} 
\end{figure} 
\begin{figure}[b]
\epsfig{file=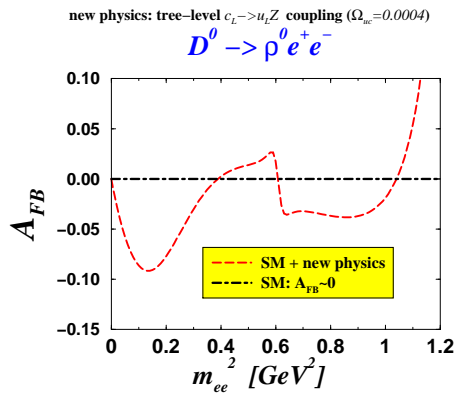,width=2.4in}
\caption{  
 The forward-backward asymmetry for $D^0\to \rho^0e^+e^-$.}
\end{figure}



We have investigated impact   of the tree-level flavor changing 
neutral transition $c \to u Z$ on the rare $D$ meson decay observables. 
However, the most suitable 
$D^+ \to \pi^+ l^+l^-$ and $D^0 \to \rho^0 l^+ l^-$ decays are found to be 
dominated by the SM long distance contributions. Only small enhancement 
of the differential mass distribution can be seen in the case of $D^+ \to \pi^+ l^+ l^-$ decay at high dilepton mass and  tiny forward backward asymmetry can be induced by new physics in $D^0 \to \rho^0 l^+ l^-$ decay.

We conclude that the new physics scenarios which contain an extra singlet heavy up-like quark,  have rather small effects on the 
charm meson observables.

\end{document}